\documentclass[12pt,a4paper,fleqn]{scrartcl}

\usepackage{graphicx} 
\usepackage{amsmath} 
\usepackage{amssymb} 
\usepackage{authblk} 
\usepackage{natbib} 

\usepackage{framed}
\usepackage{lipsum}
\usepackage{verbatim}

\title{Calculating Profits and Losses for Algorithmic Trading Strategies} 
\subtitle{A Short Guide} 
\author[1]{J.B. Glattfelder and T. Houweling}
\date{\small \today}

\begin{document}

\maketitle

\begin{abstract} \noindent
 We present a series of equations that track the total realized and unrealized profits and losses at any time, incorporating the spread. The resulting formalism is ideally suited to evaluate the performance of trading model algorithms.
\end{abstract}

\tableofcontents

\section{Introduction}

Finance and economics are characterized by a high level of formal abstraction, akin to the level of mathematization seen in physics. It is thus no surprise to encounter monographs entitled \textit{The Statistical Mechanics of Financial Markets} or \textit{Physics and Finance} \citep{voit2005statistical,ziemann2021physics}. Indeed, this analytical sophistication has sometimes been criticized for its lack of empirical justification and validation \citep{bouchaud2008economics,labini2016science}. In stark contrast, the practitioners of finance appear to face a seemingly simple challenge when assessing performance measures associated with their trading activity. After all, only basic arithmetic operations are required, alongside the calculation of percentages. As a result, these topics are mostly glossed over in standard textbooks on finance \citep{hull2000options}, where spot trading is often covered only tangentially \citep{chan2013algorithmic}, or reduced to measures relating to the net asset value and mark-to-market valuations.

Nonetheless, calculating the profits and losses (PnL) from any trading activity is a multifaceted task that can be approached in various ways, each with its own set of assumptions, advantages, and limitations. For instance, some quantitative analysts use logarithmic mid prices to compute returns \citep[p. 37]{dacorogna2001introduction}, while others utilize bid/ask prices. Choices must be made regarding the handling of fees, the utilization of compounded calculations, and whether PnL is realized or remains unrealized. In the following, we outline a straightforward methodology addressing these topics in detail.

Given a currency pair, the first currency is called the base currency ($\mathtt{b}$) and the second one is the quote currency ($\mathtt{q}$). The price at time $t_i$ is defined as $x_i$. We follow a balance-sheet bookkeeping approach. As an example, the actions of buying $u_1$ units of a currency pair at $t_1$ and then $u_2$ units at $t_2$ is represented as
\begin{table}[h!]
\centering
\begin{tabular}{c|c}
$\mathtt{b}$ & $\mathtt{q}$ \\ \hline
$\mathtt{b}_1 = u_1$ & $\mathtt{q}_1 = - x_1 u_1 $ \\
$\mathtt{b}_2 = \mathtt{b}_1 + u_2$ & $\mathtt{q}_2 = \mathtt{q}_1 - x_2 u_2$ \\
\dots & \dots
\end{tabular}
\end{table}\\ \noindent
Selling $u_i > 0$ units of a currency pair would be accounted for as $\mathtt{b}_i = \mathtt{b}_{i-1} -u_i$ and $\mathtt{q}_i = \mathtt{q}_{i-1} + u_i x_i$. In summary
\begin{subequations}
  \begin{align}
    \mathtt{b}_i  &=  \sum_{j=1}^i u_j, \\
    \mathtt{q}_i &= -\sum_{j=1}^i x_j u_j.
  \end{align} \label{eq:basic}
\end{subequations} \noindent

Note that when units are bought the trading price is the ask price and when selling it is the bid price, respectively. Thus, $x_i$ will be assumed to be the bid or ask price according to the trading context and $x_i^\prime$ will denote the opposite of the bid/ask pair and the spread is given by $s_i = |x_i - x^\prime_i|$.

Having set out the basics, the unrealized average price at time $t_i$ is given by
\begin{equation}
    \bar x_i = -\frac{\mathtt{q}_i}{\mathtt{b}_i}. \label{eq:av}
\end{equation}
Utilizing this calculated price the PnL can be evaluated at any given time.

\section{Returns}

Note that even when $\mathtt{b}_i$ goes to zero, $\mathtt{q}_i$ has the realized PnL accounted for, yielding the total calculation at all times. In general, the PnL expressed in $\mathtt{b}$ units at time $t_i$ is given by
\begin{equation}
        p^\mathtt{b}_i= 
\begin{cases}
    \mathtt{b}_i (1 - \frac{\bar x_i}{x_i^\prime}),& \text{if } \mathtt{b}_i \neq 0,\\
    -u_i (1 - \frac{\bar x_{i-1}  + \mu}{\hat x_i}),              & \text{otherwise},
\end{cases}
\label{eq:pnl}
\end{equation}
where if
\begingroup
\renewcommand{\arraystretch}{1.5}
\setlength{\tabcolsep}{12pt}
\begin{table}[h!]
\centering
\begin{tabular}{l||c|c|}
 & $\mathtt{b}_{i-1} > 0$ &  $\mathtt{b}_{i-1} < 0$ \\ \hline \hline
$\mathtt{q}_i > 0$ & $\mu = s_i$, \; $\hat x_i = x_i^\prime$ & $\mu = 0$, \; $\hat x_i = x_i$ \\ \hline
$\mathtt{q}_i < 0$ & $\mu = 0$, \; $\hat x_i = x_i$ & $\mu = -s_i$, \; $\hat x_i = x_i^\prime$ \\ \hline
\end{tabular}
\end{table} \endgroup \\  \noindent
In summary, when positions are closed ($\mathtt{b}_i = 0$) the calculation depends on the direction of the closing trade and if it yields a profit or a loss. Note that if mid prices $x_i^\dag$ are used in the computation, the expression simplifies, as $\hat x_i = x_i = x_i^\prime = x_i^\dag$ and $s_i = 0$.

To illustrate, consider the following trades for the pair SOL/USDT
\begin{table}[h!]
\centering
\begin{tabular}{l|r|r|r|r|r|r|l}
Action & $u$ & $x$ & $x^\prime$ & $\mathtt{b}$ (SOL) & $\mathtt{q}$ (USDT) & $\bar x$ & $p^{\mathtt{b}}$ (SOL) \\ \hline
Buy 5 SOL $@$170 & 5 & 170.00 & 169.75 & 5 & -850 & 170.0 & -0.007363\dots \\
Buy 10 SOL $@$175 & 5 & 175.00 & 174.75 & 15 & -2600 & $173.\overline{3}$ & \phantom{} 0.121602\dots \\
Sell 20 SOL $@$180 & -20 & 180.00 & 180.25 & -5 & -1000 & 200.0 & \phantom{} 0.547850\dots \\
Buy 5 SOL $@$160 & 5 & 160.00 & 159.75 & 0 & 200 & NaN & \phantom{} 1.25 \\
Buy 12 SOL $@$165 & 12 & 165.00 & 164.75 & 12 & -1780 & $148.\overline{3}$ & \phantom{} 1.195751\dots \\
Sell 12 SOL $@$170 & -12 & 170.00 & 170.25 & 0 & 260 & NaN & \phantom{} 1.527165\dots \\
\end{tabular}
\end{table}\\ \noindent
Observe that the profit of USDT 260 can be sold for SOL at the ask price of 170.25 resulting in a profit of SOL $1.527165\dots$, as Eq. (\ref{eq:pnl}) correctly computes.

For any currency pair, the base currency is the anchor for the PnL computation. Moreover, $p^\mathtt{b}_i$ computes the total of the PnL at time $t_i$. Thus, to retrieve the individual trade returns in base units the differences are caclculated
\begin{equation}
    \delta p^\mathtt{b}_i = p^\mathtt{b}_i  - p^\mathtt{b}_{i-i}.
\end{equation}

The PnL in quote units at time $t_i$ is given by $p^\mathtt{q}_i$. In order to evaluate this expression the opposite trade prices $x_i^\prime$ are utilized
\begin{subequations}
  \begin{align}
    p^\mathtt{q}_i  &=  p^\mathtt{b}_i x_i^\prime, \label{eq:pnlqa} \\
    \delta p^\mathtt{q}_i &= p^\mathtt{q}_i  - p^\mathtt{q}_{i-i}.
  \end{align} \label{eq:pnlq}
\end{subequations} \noindent
Note that $\delta p^\mathtt{b}_i x_i^\prime \neq \delta p^\mathtt{q}_i$.
Continuing with the example given above
\begin{table}[h!]
\centering
\begin{tabular}{l|r|l|l|r|r}
Action & $x^\prime$ & \multicolumn{1}{|r|}{$p^{\mathtt{b}}$} & \multicolumn{1}{|r|}{$\delta p^{\mathtt{b}}$} & $p^{\mathtt{q}}$ & $\delta p^{\mathtt{q}}$  \\ \hline
Buy 5 SOL $@$170 & 169.75 & -0.007363\dots  &  -0.007363\dots  & -1.2500 &  -1.25 \\
Buy 10 SOL $@$175 & 174.25 & \phantom{} 0.121602\dots &  \phantom{} 0.128966\dots & 21.2500 &  22.50 \\
Sell 20 SOL $@$180 & 180.25 & \phantom{} 0.547850\dots &  \phantom{} 0.426247\dots &  98.7500 & 77.50  \\
Buy 5 SOL $@$160 & 159.75 & \phantom{} 1.25 &  \phantom{} 0.702149\dots &  200.0000 &  101.25 \\
Buy 12 SOL $@$165 & 164.75 &  \phantom{} 1.195751\dots &   -0.054248\dots & 197.0000 &  -3.00 \\
Sell 12 SOL $@$170 & 170.25 & \phantom{} 1.527165\dots &   \phantom{} 0.331414\dots & 260.0000 &  63.00 \\
\end{tabular}
\end{table} \\ \noindent
It holds that $\sum_i \delta p^\mathtt{b}_i = 1.527165\dots$ and $\sum_i \delta p^\mathtt{q}_i = 260.0$. Observe that $56.423368\dots$ $= 0.331414\dots$ $\times 170.25 = \delta p^\mathtt{b}_6 x_6^\prime \neq \delta p^\mathtt{q}_6 = 63.0$.

As a final remark, if trading fees are charged, depending on if they are denoted in the base or quote currency, they need to be deducted from $\delta p^\mathtt{b}_i$ or $\delta p^\mathtt{q}_i$, respectively.

\section{Performance}

Assume you have an initial balance of USDT 75'000 worth SOL 500 (i.e., $x_0 = 150$) and you want to trade SOL/USDT. Note that to be able to sell any SOL you don't hold you need to short sell\footnote{This requires a margin account. Alternatively, you could also hold a balance of SOL that is hedged and sell a fraction of those assets.}. As mentioned, SOL is the anchor currency and we denote $\mathcal{B} = 500$ as the balance expressed in base units. We choose the convention setting $100 \% \equiv 1.0$.

We can now apply the PnL calculation from the given example to this context. As a result, it is possible to express the returns as percentage values $\tilde p_i$, a general measure of performance. To do so, the traded units need to be expressed as percentages $\tilde u_i$, impacting $\mathtt{b}_i$ and $\mathtt{q}_i$ in Eq. (\ref{eq:pnl}). In detail, $\tilde{{u}_i} = {u}_i / \mathcal{B}$ and thus $\mathtt{b}_i  \mapsto \widetilde{\mathtt{b}_i}$ and subsequently $\mathtt{q}_i  \mapsto \widetilde{\mathtt{q}_i}$. As a result, $\widetilde{\mathtt{b}_i}$ and $\widetilde{\mathtt{q}_i}$ are simple percentages of the balance $\mathcal{B}$. With these new inputs, the tables of the above examples can be extended
\begin{table}[ht!]
\centering
\begin{tabular}{r|r|r|r|r}
$u$ & $\mathtt{b}$ & $\tilde u$ & $\widetilde{\mathtt{b}}$ & $\widetilde{\mathtt{q}}$ \\ \hline
5 & 5 & 0.01 & 0.01 & -1.70 \\
10 & 15 & 0.02 & 0.03 & -5.20\\
-20 & -5 & -0.04 & -0.01 & 2.00\\
5 & 0 &  0.01 & 0 & 0.40 \\
12 & 12 &  0.024 & 0.024 & -3.56\\
-12 & 0 & -0.024 & 0 & 0.52\\
\end{tabular}
\end{table} \\ \noindent
Note that Eq. (\ref{eq:av}), expressing the average price computation for $\bar x_i$, is unchanged for $\widetilde{\mathtt{b}_i}$ and $\widetilde{\mathtt{q}_i}$. Applying the percentage values to Eqs. (\ref{eq:pnl}) -- (\ref{eq:pnlq}) yields the performance values $\tilde p_i = \tilde p_i (u_i, \mathcal{B}, x_i, x_i^\prime, \bar x_i, \hat x_i, \mu)$. Following the example
\begin{table}[ht!]
\centering
\begin{tabular}{l|l}
 \multicolumn{1}{r|}{$\tilde p_i$} &  \multicolumn{1}{|r}{$\delta \tilde p_i$} \\ \hline
-0.00001472\dots & -0.00001472\dots \\
\phantom{-}0.00024320\dots & \phantom{-}0.00025793\dots \\
\phantom{-}0.00109570\dots & \phantom{-}0.00085249\dots \\
\phantom{-}0.00250000\dots & \phantom{-}0.00140429\dots \\
\phantom{-}0.00239150\dots & -0.00010849\dots \\
\phantom{-}0.00305433\dots & \phantom{-}0.00066282\dots
\end{tabular}
\end{table} \\ \noindent
In summary, given $u_i$, $\mathcal{B}$ and the bid/ask prices, $\tilde p_i$ can be derived. Now, the total PnL in base and quote units can be easily computed as
\begin{subequations}
  \begin{align}
    p^\mathtt{b}_i  &=  \tilde p_i \mathcal{B}, \\
    p^\mathtt{q}_i &= \tilde p_i \mathcal{B} x_i^\prime.
  \end{align} \label{eq:pnlp}
\end{subequations}
The compounded return is given by $\hat p_n = (\delta \tilde p_1 + 1) \times \dots \times (\delta \tilde p_n + 1) -1$. 

In the example, $p^\mathtt{b}_6 = 0.00305433\dots \times 500 = 1.527165\dots$ and $p^\mathtt{q}_6 = 1.527165\dots \times 170.25 = 260$. Also observe that $\widetilde{\mathtt{q}_6} / x_6^\prime = 0.52 / 170.25 = 0.00305433\dots = p^\mathtt{b}_6$ and that $\widetilde{\mathtt{q}_6} \mathcal{B} x_6^\prime = \mathtt{q}_6 = 260 = p^\mathtt{q}_6$. Finally, $\tilde p_6 = \sum_j \delta \tilde p_j =  0.00305433\dots$

\section{Wealth}

Now assume you have an initial base balance of $\mathcal{B} = 500$ and in addition an initial quote balance of $\mathcal{Q} = 75'000$. Our total wealth $\mathcal{W}$ is defined as the sum of the two balances, split up equally\footnote{The PnL computations of this section do not depend on this assumption.} at $t_0$ (at a conversion rate of $x_0 = 150$). Note that both balances can be evaluated in the opposite currency $\mathcal{B} \longleftrightarrow \mathcal{B}^{\,\mathtt{q}}$ and $\mathcal{Q} \longleftrightarrow  \mathcal{Q}^\mathtt{b}$ yielding $\mathcal{W}$ in base or quote units. In detail, the initial wealth evaluated at the price $x_i$ is
\begin{subequations}
  \begin{align}
    \mathcal{\overline W}^\mathtt{b}_i  &:=  \mathcal{B} +\mathcal{Q}^{\mathtt{b}} :=  \mathcal{B} + \frac{\mathcal{Q}}{x_i^\prime}, \\
     \mathcal{\overline  W}^\mathtt{q}_i  &:=  \mathcal{Q} + \mathcal{B}^{\, \mathtt{q}} :=  \mathcal{Q} +  x_i^\prime \mathcal{B}. 
  \end{align} \label{eq:wbar}
\end{subequations}
For the given example, $\mathcal{\overline W}^\mathtt{b}_0 = 1'000$ and $\mathcal{\overline W}^\mathtt{q}_0 = 75'000$. The evolution of the wealth will be impacted by the trading activity and the price changes. As a result, all $u_i$ and $-u_i x_i$ affect the base and quote balances
\begin{subequations}
  \begin{align}
    \mathcal{W}^\mathtt{b}_i  &=  \mathcal{B} +\sum_{j=1}^{i} u_j + \frac{1}{x_i^\prime} \left( \mathcal{Q} - \sum_{j=1}^{i} u_j x_j \right) =  \mathcal{B} + \mathtt{b}_i + \frac{1}{x_i^\prime} \left( \mathcal{Q} +  \mathtt{q}_i \right) , \\
      \mathcal{W}^\mathtt{q}_i  &= x_i^\prime \left( \mathcal{B} + \sum_{j=1}^{i} u_j \right) +  \mathcal{Q} - \sum_{j=1}^{i} u_j x_j  =  x_i^\prime ( \mathcal{B} + \mathtt{b}_i ) + \mathcal{Q} +  \mathtt{q}_i .
  \end{align} \label{eq:w}
\end{subequations}
In simple terms, $\mathcal{W}^\mathtt{b}_i =\mathcal{W}^\mathtt{q}_i / x_i^\prime$.

To evaluate the PnL at time $t_i$ we compare $\mathcal{W}^{\mathtt{b}, \mathtt{q}}_i$ to $\mathcal{\overline W}^{\mathtt{b}, \mathtt{q}}_i$, the benchmark portfolio representing a no-trading strategy. In detail
\begin{subequations}
  \begin{align}
    p^\mathtt{b}_i &= \mathcal{W}^\mathtt{b}_i  - \mathcal{\overline W}^\mathtt{b}_i = \mathtt{b}_i + \frac{\mathtt{q}_i}{x_i^\prime}, \label{eq:pnlfa} \\
     p^\mathtt{q}_i &= \mathcal{W}^\mathtt{q}_i  - \mathcal{\overline W}^\mathtt{q}_i = x_i^\prime \mathtt{b}_i + \mathtt{q}_i. \label{eq:pnlfb}
  \end{align}  \label{eq:pnlf}
\end{subequations}
Observe that, similarly to Eq. (\ref{eq:pnl}), the bid and ask prices need to be properly accounted for when $\mathtt{b}_i \to 0$. Recall that $x_i$ is the bid or ask price of the trade $u_i$ and $x_i^\prime$ refers to the opposite. Eqs. (\ref{eq:wbar}), (\ref{eq:w}), and (\ref{eq:pnlf}) assume $\mathtt{b}_i \neq 0$, utilizing $x_i^\prime$ in the conversions. However, for $\mathtt{b}_i = 0$
\begin{equation}
        x_i^\prime \to  x_i, \quad \text{if} \quad (\mathtt{b}_{i-1} > 0 \, \,  \land  \, \,\mathtt{q}_{i} > 0)  \,  \, \lor  \,  \,  (\mathtt{b}_{i-1} < 0 \,  \, \land \,  \, \mathtt{q}_{i} < 0). 
\end{equation}
Finally, to conclude with the example \\
\begin{table}[ht!]
\centering
\begin{tabular}{r|r|r||r|r|r}
$\overline{\mathcal{W}}^\mathtt{b}$ & $\mathcal{W}^\mathtt{b}$ & $p^\mathtt{b}$  & 
$\mathcal{\overline W}^\mathtt{q}$ & $\mathcal{W}^\mathtt{q}$ &  $p^\mathtt{q}$   \\ \hline
941.8262\dots & 941.8188\dots & -0.007363\dots & 159'875 & 159'873.75 & -1.25 \\
929.1845\dots & 929.3061\dots &  0.121602\dots & 162'375 & 162'396.25 & 21.25 \\
916.0887\dots & 916.6366\dots & 0.547850\dots & 165'125 & 165'223.75 &  98.75 \\
968.7500\dots & 970.0000\dots &1.25\phantom{0000\dots} & 155'000 & 155'200.00 & 200.00 \\
955.2352\dots & 956.4309\dots & 1.195751\dots & 157'375 & 157'572.00 & 197.00 \\
940.5286\dots & 942.0558\dots  & 1.527165\dots & 160'125 & 160'385.00 & 260.00 \\
\end{tabular}
\end{table} \\ \noindent
Depending on the currency chosen to express the wealth in, we see an increase or decrease in the final value due to the underlying price evolution. Thus, either, in SOL, $1'000 \to 942.0558\dotsc$ or, in USDT, $150'000 \to 160'385$. In general, if $\nu^\mathtt{b}$ is the difference between the final and the initial base wealth, then the corresponding change in quote wealth is given by $\nu^\mathtt{q} = -x_i^\prime \nu^\mathtt{b}$ and it is a preference which perspective to adopt. Regardless of this choice, the trading profit is $p^\mathtt{b}_6 = 1.5271\dots$ or $p^\mathtt{q}_6 = 260$, as can be either computed from Eqs. (\ref{eq:pnl}) and (\ref{eq:pnlqa}), (\ref{eq:pnlp}), or (\ref{eq:pnlf}). In closing, the simplest computation of the PnL resulting from trading is done utilizing Eqs. (\ref{eq:basic}) and (\ref{eq:pnlf}).


\bibliographystyle{apalike} 
\bibliography{my} 

\end{document}